\documentclass[conference]{IEEEtran}
\IEEEoverridecommandlockouts
\usepackage{cite}
\usepackage{amsmath,amssymb,amsfonts}
\usepackage{algorithmic}
\usepackage{graphicx}
\usepackage{subfigure}
\usepackage{textcomp}
\usepackage{siunitx}
\usepackage{float}
\usepackage{threeparttable} 
\usepackage{soul}
\usepackage{hyperref}
\usepackage[table,xcdraw]{xcolor}

\usepackage{multirow}
\usepackage{multicol}
\usepackage{threeparttable}
\usepackage{breqn}
\usepackage{tikz}

\def\BibTeX{{\rm B\kern-.05em{\sc i\kern-.025em b}\kern-.08em
    T\kern-.1667em\lower.7ex\hbox{E}\kern-.125emX}}

\makeatletter
\DeclareRobustCommand{\oplustimes}{%
  \mathbin{\mathpalette\o@plus@times\relax}%
}
\newcommand{\o@plus@times}[2]{%
  \ooalign{$\m@th#1\oplus$\cr$\m@th#1\otimes$\cr}%
}
\makeatother

\begin{document} 
\title{Scaling Analog Photonic Accelerators for Byte-Size, Integer General Matrix Multiply (GEMM) Kernels}

\author{\IEEEauthorblockN{ Oluwaseun Adewunmi Alo}
\IEEEauthorblockA{\textit{Electrical and Computer Engineering} \\
\textit{University of Kentucky, USA}\\
seun.alo@uky.edu}

\and
\IEEEauthorblockN{Sairam Sri Vatsavai}
\IEEEauthorblockA{\textit{Electrical and Computer Engineering} \\
\textit{University of Kentucky, USA}\\
sairam\_srivatsavai@uky.edu}

\and
\IEEEauthorblockN{Ishan Thakkar}
\IEEEauthorblockA{\textit{Electrical and Computer Engineering} \\
\textit{University of Kentucky, USA}\\
igthakkar@uky.edu}
}

\maketitle

\begin{abstract}

Deep Neural Networks (DNNs) predominantly rely on General Matrix Multiply (GEMM) kernels, which are often accelerated using specialized hardware architectures. Recently, analog photonic GEMM accelerators have emerged as a promising alternative, offering vastly superior speed and energy efficiency compared to traditional electronic accelerators. However, these photonic cannot support wider than 4-bit integer operands due to their inherent trade-offs between analog dynamic range and parallelism. This is often inadequate for DNN training as at least 8-bit wide operands are deemed necessary to prevent significant accuracy drops. To address these limitations, we introduce a scalable photonic GEMM accelerator named SPOGA. SPOGA utilizes enhanced features such as analog summation of homodyne optical signals and in-transduction positional weighting of operands. By employing an extended optical-analog dataflow that minimizes overheads associated with bit-sliced integer arithmetic, SPOGA supports byte-size integer GEMM kernels, achieving significant improvements in throughput, latency, and energy efficiency. Specifically, SPOGA demonstrates up to 14.4$\times$, 2$\times$, and 28.5$\times$ improvements in frames-per-second (FPS), FPS/Watt, and FPS/Watt/mm² respectively, compared to existing state-of-the-art photonic solutions.

\end{abstract}


\section{Introduction}

Deep Neural Networks (DNNs) can process complex data and perform tasks such as image recognition, natural language processing, and speech recognition with remarkable accuracy. For processing DNNs, GEMM functions play a crucial role, particularly in various linear layers of DNNs such as fully connected, convolutional, and self/cross attention layers \cite{b19}. 

Some of these DNN layers do not directly employ GEMMs but their comprising computational kernels are often transformed into GEMM functions for efficient hardware-based acceleration. For instance, convolution layers are often converted into input and Toeplitz matrices using Im2Col operations to enable GEMM functions between these matrices \cite{b7}. DNNs also employ non-linear functions, along with linear functions such as GEMM functions. However, GEMM functions often comprise about 70\% of the execution cycles while training DNNs \cite{b19}, making them the key workload in all DNNs. 

While GEMM functions continue to remain the favorite abstraction to which the tensors during the forward and backward passes of DNNs are unrolled, the performance and energy efficiency demands for processing the GEMM functions of modern DNNs have grown towards becoming unsustainable over the past decade. This is due to the number of trainable parameters of DNNs growing from thousands to more than hundreds of billions, and the speed and energy consumption of traditional hardware architectures, such as Graphics Processing Units (GPUs) \cite{b19}, Tensor Processing Units (TPUs) \cite{b25}, and Application-Specific Integrated Circuits (ASICs) \cite{b20} based designs, not scaling sufficiently to keep up. As a result, to meet these demands, hardware architects dwell in search of ultrafast and extremely energy-efficient architectures to accelerate GEMM functions.  

Fortunately, this search has led to the recent demonstrations of analog photonic accelerators for GEMM functions \cite{b2,b3,b5,b6,b8,b9}. It has been shown that analog photonic accelerators can achieve two to three orders of magnitude higher processing speed and energy efficiency for processing GEMM functions compared to other digital/analog electronic accelerators \cite{b1,b2,b3,b4,b6,b9,b22}. These benefits of analog photonic accelerators are attributed to their massive, multi-dimensional parallelism along with the ultra-low-dissipation, impedenceless, and high-speed dynamics. 

Despite these advantages, however, analog photonic accelerators for GEMM functions face daunting challenges due to the strong trade-offs among their achievable multi-dimensional parallelism, operating speeds, and operand bit-widths. It has been shown that, due to these trade-offs, the achievable operand bit width for analog multiplications in analog photonic accelerators for GEMM functions remains constricted to 4-bit with fixed-point precision at 1 gigasamples (GS)/sec speed and 44$\times$44 in-parallel multiplications per GEMM core \cite{b1,b2}. As the operand bit-width increases to fixed-point 8-bit, the achievable parallelism diminishes to only 1 multiplication per core. This occurs because a major portion of the already tight optical power budget is utilized to support the large dynamic range required for 8-bit precision (a total of 2$^8$=256 analog optical power levels are required for 8-bit operands compared to only 2$^4$=16 levels required for 4-bit operands), leaving an insufficient portion of the optical power budget in support of high optical parallelism \cite{b2}. 

Although 4-bit fixed-point precision is touted to be sufficient for many DNN inference applications accelerated on the edge devices, it has been shown that a minimum of 8-bit fixed-point operands are required for multiplications with at least 16-bit precision required for intermediate accumulations (before their rounding to 8-bit precision) during DNN training to keep the drop in the accuracy tolerable while simultaneously achieving high energy efficiency \cite{b26,b27}. Thus, there is a crucial need to scale analog photonic architectures to accelerate GEMM functions with byte-size integer operands. 

To meet this requirement, the analog photonic accelerators in the literature often utilize bit-sliced operands \cite{b3,b9}, wherein the INT8 operands are split into two concatenated INT4 slices. Consequently, the processing of every GEMM function between matrices comprising INT8 operands is decomposed into four GEMM functions each between matrices comprising INT4 operands. Each INT4 GEMM function is implemented on a dedicated analog photonic core to produce an intermediate output matrix. All four intermediate matrices, thus produced, are then post-processed to generate the final result of the INT8 GEMM function. Such processing using INT4 slices of INT8 operands incurs excessive area, latency, and power overheads for analog-to-digital conversions, memory storage and access of intermediate matrices, and digital electronic post-processing of intermediate results. These overheads diminish the inherent throughput and energy efficiency benefits of employing analog photonic computing. 

To overcome these shortcomings, we present a \textbf{\underline{S}}calable \textbf{\underline{P}}hot\underline{\textbf{O}}nic \textbf{\underline{G}}EMM \textbf{\underline{A}}ccelerator (SPOGA) for byte-size integer GEMM kernels in this paper. SPOGA introduces these two enhanced features in the employed electro-photonic circuits: (1) analog summation of homodyne optical signals by a time-integrating optical-to-electrical transduction circuit through incoherent superposition and charge accumulation, and (2) in-transduction weighting of accumulated analog results. These features extend the mixed optical/analog dataflow in SPOGA for the temporary storage and post-processing of intermediate matrices generated from INT4-sliced GEMM functions. This in turn eliminates the above-mentioned overheads of processing bit-sliced integer operands, rendering substantial throughput, latency, and energy-efficiency benefits to SPOGA for accelerating byte-size integer GEMM functions. 

The rest of the paper is organized as follows. Section II discusses related work and provides background and motivation. Section III provides an overview of the SPOGA architecture and describes its key enhanced features. The scalability analysis and the results of benchmark-drivel system-level evaluations obtained for four modern DNNs are presented in Section IV. Section V concludes this paper.

\section{Background and Motivation}





\begin{figure*}
    \centering
    \subfigure[Illustration of a General Matrix Multiplication (GEMM) operation.]
    {
        \includegraphics[width=30pc]{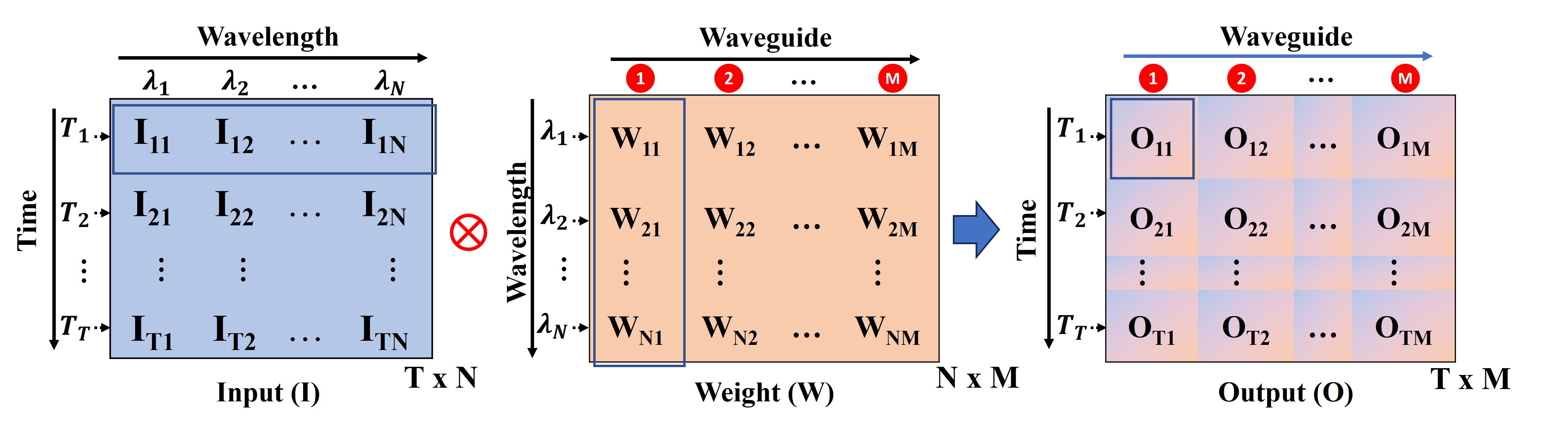}
        \label{fig:gemm_op}
    }
    \subfigure[Mapping of GEMM operation from (a) to MAW and AMW types of GEMM core organisations from prior works.]
    {
        \includegraphics[width=38pc]{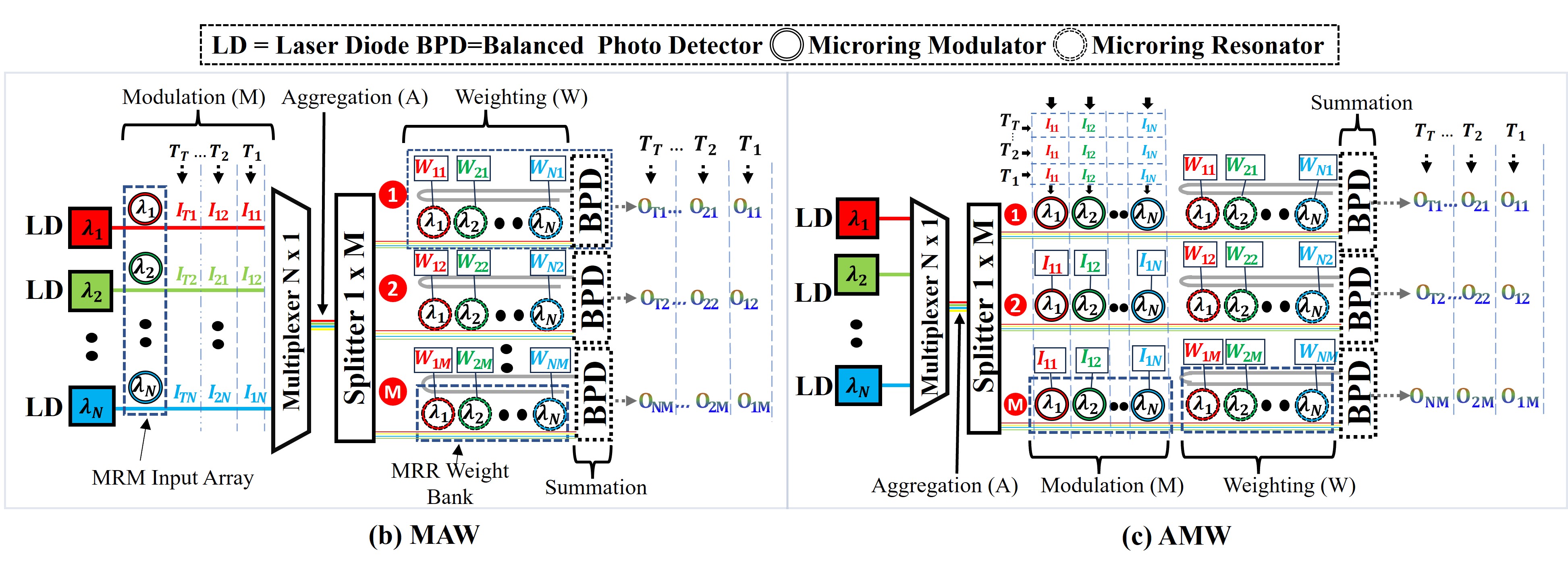}
        \label{fig:amw_maw}
    }
    \caption{Illustration of a General Matrix Multiplication (GEMM) operation and its mapping on photonic GEMM cores.}
    \label{fig:Gemm}
\end{figure*}

\subsection{Related Work on Analog Photonic GEMM Accelerators} \label{sec2c}
It has been shown that analog photonic accelerators can achieve two to three orders of magnitude higher processing speed/throughput and energy efficiency for processing GEMM functions compared to other digital/analog electronic accelerators \cite{b24}. The higher processing speed and throughput of analog photonic accelerators are attributed to their massive, multi-dimensional parallelism \cite{b24}. Similarly, their higher energy efficiency is attributed to their ultra-low-dissipation, impedanceless, and high-speed dynamics. The analog photonic GEMM accelerators from the literature can be broadly categorized into two main types: coherent and non-coherent architectures. Incoherent GEMM accelerators \cite{b2,b3,b6,b9} modulate optical signal power to represent tensor values, whereas coherent GEMM accelerators \cite{b5} utilize optical field amplitude and phase. Our study focuses on incoherent architectures because of their demonstrated scalability and performance advantages over their coherent counterparts \cite{b5}. Non-coherent architectures utilize parallel wavelengths and waveguides, with parameters encoded within optical wavelength signals' amplitudes using devices like microring resonators (MRRs).

The MRR-based incoherent photonic accelerators from the literature often employ multiple GEMM cores that operate concurrently. These GEMM cores are primarily responsible for executing Vector Dot Product (VDP) operations that are derived by decomposing the target GEMM functions. Every GEMM core employs a total of N laser diodes that generate N optical wavelength channels \(\lambda_{1}\ - \lambda_{N}\)\. These optical wavelength channels are manipulated in five different ways by five different blocks of photonic devices and circuits to implement parallel VDP operations. These blocks include (1) A splitting block that employs a series of splitters for splitting (copying) \textit{N} optical signals in \textit{M} waveguides to achieve a fan-out degree of \textit{M} per optical signal; (2) An aggregation block that employs multiplexers for aggregation (multiplexing) of \textit{N} optical signals per waveguide to achieve a fan-in degree of \textit{N} per waveguide; (3) A modulation block that employs one or more MRR modulator (MRM) arrays for modulation of optical signals to imprint input values onto them; (4) A weighting block that employs multiple parallel MRR weighting banks for weighting of modulated optical signals to achieve analog input-weight products; (5) A summation block that comprises a total of \textit{M} balance photodetectors (BPDs) each coupled with a time-integrating or trans-impedance receiver to perform analog summation of optical signals through incoherent superposition and charge accumulation. The specific order in which these blocks are arranged distinguishes various prior GEMM core designs. These designs can be broadly categorized into MAW (Modulation-Aggregation-Weighting) \cite{b8}, AMW (Aggregation-Modulation-Weighting) \cite{b9}, and MWA (Modulation-Weighting-Aggregation). For each GEMM core, the defined sequence of these signal manipulations plays a pivotal role in the efficiency and performance of the core, ultimately shaping the core's computational capabilities. The reader is directed to \cite{b23} for more details.

While these architectures excel in handling GEMM operations, recent advancements have also shown all-optical and opto-electronic methods for implementing non-GEMM operations, such as activation operations, as well \cite{b28}, although we only focus on GEMM operations in this work.

\subsection{Mapping of GEMM Functions on Photonic Accelerators} 
Both electronic and photonic GEMM accelerators support temporal, spatial, and mixed spatio-temporal mapping of GEMM functions onto GEMM cores. However, photonic accelerators provide additional advantages in spatial mapping, offering two distinct approaches: mapping by waveguide and by wavelength. This increased degree of freedom in spatial mapping provides a notable advantage over their electronic counterparts. For instance, Fig. \ref{fig:Gemm} illustrates how a GEMM function is typically mapped onto the MAW and AMW categories of GEMM cores. Each GEMM core design in Fig. \ref{fig:amw_maw} is shown to have a total of \textit{N} wavelength channels and \textit{M} waveguides. From Fig. \ref{fig:gemm_op}, the input matrix $\mathbf{I}$ (of size \textit{T}$\times$\textit{N}) is mapped spatially (across \textit{N} wavelengths) and temporally (across \textit{T} time steps), whereas the weight matrix $\mathbf{W}$ (of size \textit{N}$\times$\textit{M}) is mapped spatially across \textit{N} wavelengths and \textit{M} waveguides. As a result, during each time step, each of the total \textit{M} BPDs produces the sum of a total of \textit{N} products. Therefore, since there are a total of \textit{M} BPDs in each GEMM core design, a total of \textit{M} sums of products (\textit{M} dot products) are produced in parallel in each GEMM core.

\subsection{Bit-Slicing for GEMM Functions with Byte-Size Operands}
\begin{figure*}
    \centering
    \subfigure[The method used in prior works.]
    {
        \includegraphics[width=30pc]{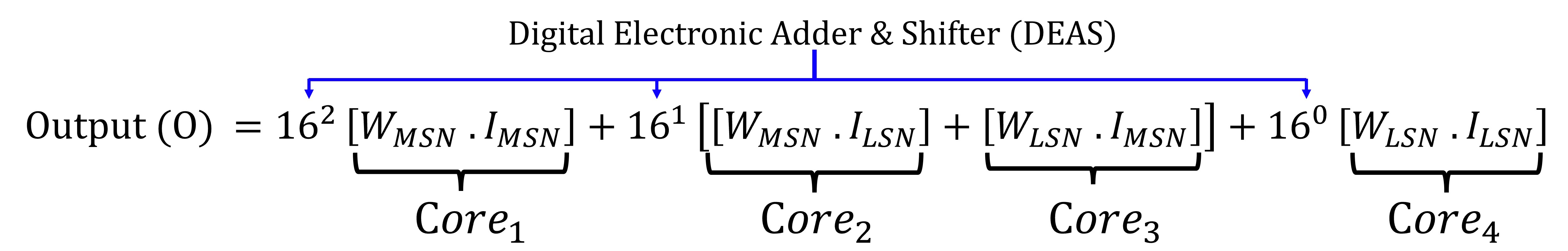}
        \label{fig:weight_ass_old}
    }
    \hfill
    \subfigure[The method used in SPOGA, at matrix-level abstraction. All the 'I's and 'W's are matrices.]
    {
        \includegraphics[width=30pc]{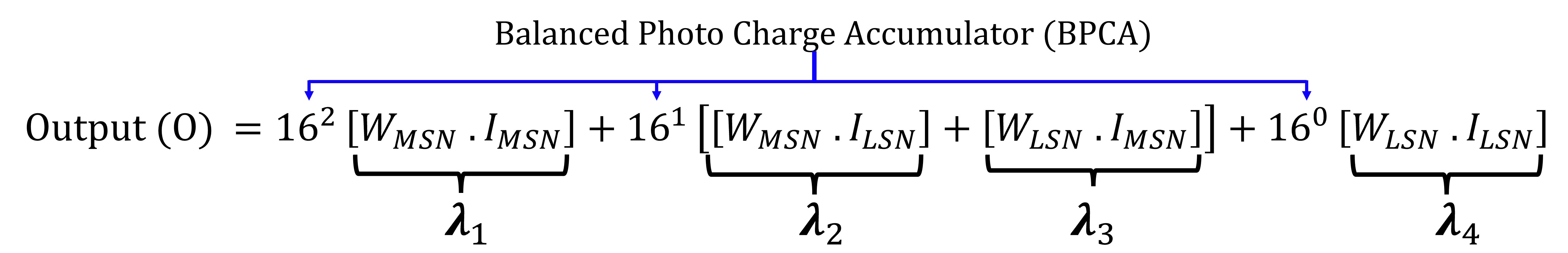}
        \label{fig:weight_ass_new}
    }
    \hfill
    \subfigure[The method used in SPOGA, at dot product level abstraction. All the 'I's and 'W's are INT4 values.]
    {
        \centering
        \includegraphics[width=\textwidth]{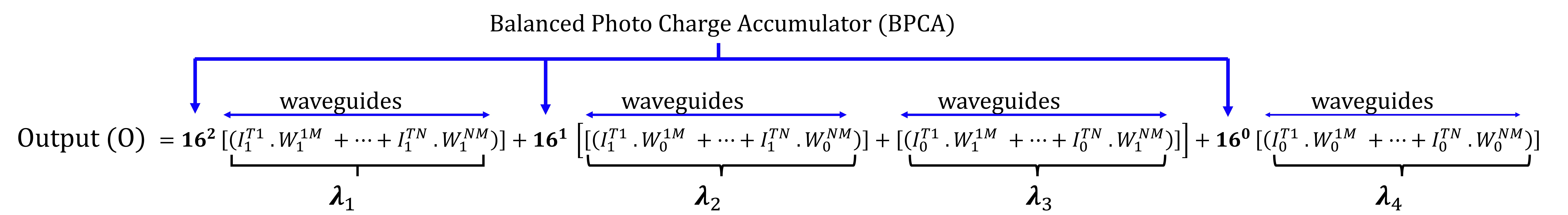}
        \label{fig:maptohw}
    }

    \caption{Different methods of implementing and mapping a GEMM function for hardware acceleration using bit-sliced integer arithmetic.}
    \label{fig:bit_slice}
\end{figure*}

S. S Vatsavai et al. in \cite{b2} inferred that photonic GEMM accelerator cores from the literature cannot support wider than 4-bit integer operands while simultaneously supporting expected processing parallelism (i.e., numbers of parallel wavelength channels and waveguides). This is because these GEMM cores exhibit a very strong trade-off between the achievable operand bitwidth and parallelism. To address this shortcoming, the analog photonic accelerators from prior works often utilize bit-sliced operands, wherein the INT8 operands are split into two concatenated INT4 slices: one slice comprising the Most Significant Nibble (MSN) and the other slice comprising the Least Significant Nibble (LSN). Consequently, the input matrix $\mathbf{I}$ and weight matrix $\mathbf{W}$ from Fig. \ref{fig:Gemm} are decomposed into two INT4 matrices each, $\mathbf{I}$$_{MSN}$/$\mathbf{I}$$_{LSN}$ and $\mathbf{W}$$_{MSN}$/$\mathbf{W}$$_{LSN}$, respectively. As a result, the processing of a GEMM function between the input and weight matrices comprising INT8 values is decomposed into four GEMM functions, each function between LSN/MSN matrices comprising INT4 values, as shown in Fig. \ref{fig:weight_ass_old}. Each INT4 GEMM function is implemented on a dedicated analog photonic GEMM core to produce an intermediate output matrix. All four intermediate matrices, thus produced, are then post-processed using the Digital Electronic Shifter and Adder (DEAS) block (Fig. \ref{fig:weight_ass_old}). This postprocessing using DEAS is needed to multiply the individual values of each intermediate result matrix with the weight of the corresponding radix positions (see 16$^0$, 16$^1$, and 16$^2$ as the radix position weights of the intermediate result matrices in Fig. \ref{fig:weight_ass_old}). This post-processing generates the final output matrix (Fig. \ref{fig:weight_ass_old}). 

\subsection{Motivation}
In the bit-slicing-based approach, the DEAS-based post-processing of intermediate result matrices incurs very high overheads. This is because the intermediate result matrices are generated in the optical analog format by the analog photonic GEMM cores; therefore, these matrices have to be converted to the electronic digital format first. This step requires high-speed optical-to-electrical and analog-to-digital converters (ADCs) because at each BPD one result needs to be converted to the electrical and then digital format every time step of 1 nanosecond or sub-nanosecond (corresponding to $\geq$1 GS/sec speed). Such high-speed optical-to-electrical converters and ADCs consume non-trivial amounts of power/energy and area. After these digital conversions using ADCs, the intermediate matrices then need to be stored in digital memory and accessed from the memory for their subsequent DEAS-based processing. This consumes additional energy in addition to the extra area and latency overheads of DEAS units. Overall, these overheads diminish the inherent throughput and energy efficiency benefits of employing analog photonic computing.

\section{SPOGA Architecture: Overview}

The main idea of our SPOGA architecture is to extend the optical-analog dataflow inside a GEMM core to implement the radix-position-based weighting and addition of the intermediate result matrices opto-electronically directly in the analog format during the optical-to-electrical transduction, without requiring to convert the intermediate results matrices in the digital format nor having to employ sluggish DEAS circuits. SPOGA achieves this by employing a unique GEMM core architecture that comprises a total of 16 dot product units (DPU). Each DPU is capable of performing a dot product on large input and weight vectors comprising INT8 values, with up to 249 INT8 values per vector (more on this size scalability analysis is discussed in Section \ref{sec:scalability}). Each DPU handles the radix-position-based weighting and addition of the INT4-operands-based intermediate results in the optical domain. For that, it assigns a common wavelength channel to all bit-sliced (INT4) multiplication operations that have the same radix position weight (see Figs. \ref{fig:weight_ass_new} and \ref{fig:maptohw}). This arrangement provides several benefits compared to the bit-sliced processing approach illustrated in Fig. \ref{fig:weight_ass_old}, as will be discussed in Section \ref{sec:benefits}. 

\subsection{Structure and Operation of SPOGA DPU}
Fig. \ref{fig:spoga} illustrates the structure of a SPOGA DPU. A SPOGA DPU comprises three unique types of photonic circuits: (1) multiple Optical Analog Multiplier Ensembles (OAMEs), (2) radix-position-weight-aware aggregation lanes, and (3) one Positional Weighting and Accumulation Block (PWAB) for radix-position-based weighting and accumulation/addition of intermediate results.
\subsubsection{Optical Analog Multiplier Ensemble (OAME)} 
\begin{figure*}
\centerline{\includegraphics[width=\linewidth]{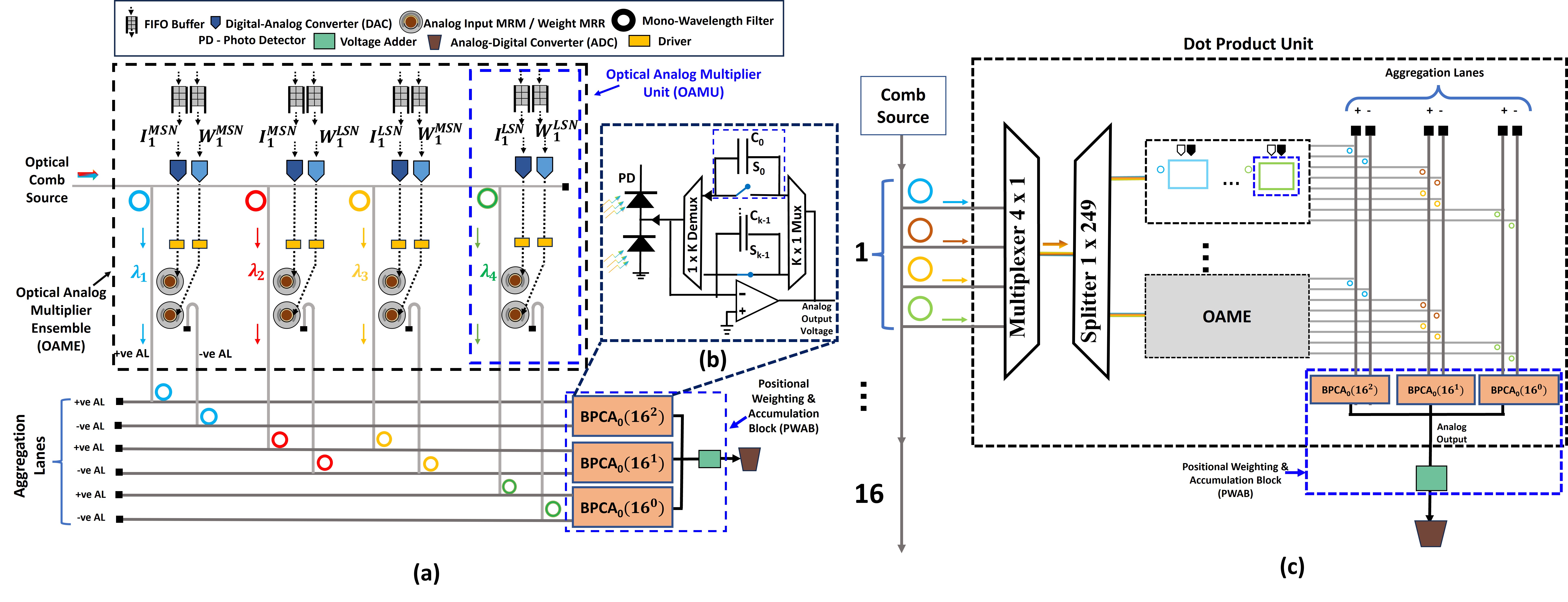}}
\caption{SPOGA Architecture Overview. (a) An Optical Analog Multiplier Ensemble (OAME), a component that composes a SPOGA GEMM core comprising Dot Product Units (DPUs) in (c). (b) A Balanced Photo Charge Accumulator (BPCA) that composes the Positional Weighting and Accumulation Block (PWAB) in (a) and (c).}
\label{fig:spoga}
\end{figure*}

From Fig. \ref{fig:spoga}(a), each OAME performs a multiplication between two INT8 operands, i.e., INT8 input I$_1$ and INT8 weight W$_1$, in a bit-sliced manner. Since each operand is sliced into INT4 MSN and LSN (e.g., I$_1$ is sliced into I$_1^{MSN}$ and I$_1^{LSN}$, and W$_1$ sliced into W$_1^{MSN}$ and W$_1^{LSN}$) the OAME performs a total of four INT4 multiplications, i.e., (i) I$_1^{MSN}$$\times$W$_1^{MSN}$, (ii) I$_1^{MSN}$$\times$W$_1^{LSN}$, (iii) I$_1^{LSN}$$\times$W$_1^{MSN}$, and (iv) I$_1^{LSN}$$\times$W$_1^{LSN}$. For that, the OAME employs four parallel optical analog multiplier units (OAMUs) (see Fig. \ref{fig:spoga}(a)). These four OAMUs are assigned four dedicated wavelength channels ($\lambda_1$, $\lambda_2$, $\lambda_3$, and $\lambda_4$), with one wavelength channel assigned per OAMU. As a result, the result of the INT4 multiplication I$_1^{MSN}$$\times$W$_1^{MSN}$ is carried on a dedicated wavelength $\lambda_1$, and so forth for the remaining four INT4 multiplications as well (see Fig. \ref{fig:spoga}(a)). The four intermediate result values that are produced by the OAME emerge from the OAME on four different wavelength channels. These values are unweighted as they emerge, and they need to undergo radix-position-based weighting and consequent accumulation/addition, to produce the final output. For that, SPOGA DPU employs aggregation lanes and a PWAB, as discussed next. 

\subsubsection{Aggregation Lanes} Each intermediate result emanating on a wavelength from an OAME needs to undergo a different radix-position-based weighting. For instance, the weighting associated with the result carried by $\lambda_1$ should be 16$^2$. This is because $\lambda_1$ corresponds to an OAMU that produces a multiplication result between two MSNs. Each MSN of an INT8 operand has a radix-positional weight of 16$^1$. As a result, a multiplication of two MSNs should have a positional weight of 16$^2$, making the positional weight of $\lambda_1$ to be 16$^2$. Similarly, since there is one MSN multiplied with one LSN in the OAMUs corresponding to wavelengths $\lambda_2$ and $\lambda_3$, the associated weight in this case should be 16$^1$ for both wavelengths. Along the same lines, since there is one LSN multiplied with another LSN in the OAMU corresponding to wavelengths $\lambda_4$, the weight associated with $\lambda_4$ should be 16$^0$. All of these positional weights are applied to the unweighted intermediate results in a PWAB (Fig. \ref{fig:spoga}(a)). Therefore, all the results-carrying wavelengths emanating from an OAME needs to be sent to the PWAB. This is achieved using the aggregation lanes (Fig. \ref{fig:spoga}(a)). There are a total of three sets of aggregation lanes. Each set corresponds to a specific positional weight. Wavelengths $\lambda_4$ and $\lambda_1$ are propagated to the PWAB in the aggregation lane sets corresponding to the positional weights of 16$^0$ and 16$^2$ respectively. In contrast, wavelengths $\lambda_2$ and $\lambda_3$ are aggregated (multiplexed) in the aggregation lane set corresponding to positional weight 16$^1$. Each lane set has one positive (+ve) lane and one negative (-ve) lane. A wavelength takes either +ve lane or -ve lane based on the sign of the result values carried onto the wavelength. It is noteworthy that a SPOGA DPU comprises several (up to 249) OAMEs. The three aggregation lane sets discussed above are shared among all OAMEs of a DPU (Fig. \ref{fig:spoga}(c)) so that the unweighted-results-carrying signals of carrier wavelength $\lambda_1$ emanating from all OAMEs are multiplexed into the lane sets corresponding to positional weight 16$^2$, and so forth. These shared aggregation lane sets take the multiplexed wavelength signals to the PWAB.

\subsubsection{Positional Weighting and Accumulation Block (PWAB)} A PWAB comprises three parallel balance photo-charge accumulators (BPCAs) \cite{b1,b22, b23} followed by an analog voltage adder and an ADC. Each BPCA comprises a BPD connected to a time-integrating receiver with a bank of selectable accumulation capacitors \cite{b2}. The three parallel BPCAs correspond to three positional weight values 16$^2$, 16$^1$, and 16$^0$. During a time step of interest, each BPCA acts as an optical-to-electrical transducer and spatial accumulator to produce an analog output voltage that is proportional to the sum of optical signals (product values) provided as input during the time step. Thus, it produces a dot product result in each time step. The BPCA design from our prior works \cite{b1,b2} only sums heterodyne optical signals (signals of different optical carrier wavelengths). In contrast, in this work, we re-purpose the BPCA to perform the summation of homodyne optical signals. This is crucial because each BPCA in Fig. \ref{fig:spoga} connects with exactly one aggregate lane set that carries unweighted results from multiple OAMEs on optical signals of the same carrier wavelength. Hence, summing these homodyne optical signals together in turn accumulates the unweighted results together into an intermediate partial result. Thus, the use of three BPCAs produces three positionally unweighted partial results. In addition, we added another feature to our designed BPCA. We enabled each BPCA to select one specific accumulation capacitor with the capacitance of 16$^{-2}$$\times$C$_0$, 16$^{-1}$$\times$C$_0$, or C$_0$ to scale its output analog voltage by the factor of 16$^2$, 16$^1$, or 16$^0$, respectively, for the same input. This enables each BPCA to multiply its partial result with its corresponding positional weight (i.e., 16$^2$, 16$^1$, or 16$^0$). This way, the three BPCAs produce three positionally weighted partial results in the analog format. These weighted partial results are then summed together using an analog voltage adder, the result of which is converted into the final digital result using an ADC (Fig. \ref{fig:spoga}(c)). This final digital result is the dot product output. Thus, the PWAB block enables the positional weighting and accumulation of intermediate partial results during the optical-to-electrical traduction of input optical signals.

\subsection{Benefits of Extending Optical-Analog Dataflow in SPOGA}\label{sec:benefits}
Due to the unique design of a SPOGA DPU, the generation of one dot product output in the digital format requires three optical-to-electrical conversions (of three partial results) and one analog-to-digital conversion. It also does not require intermediate memory storage or post-processing using DEAS. This incurs much less overhead than the approach from Fig. \ref{fig:weight_ass_old}, which requires a total of four optical-to-electrical conversions, four analog-to-digital conversions, intermediate memory storage, and post-processing using DEAS. The notable reduction in the overheads translates to substantially better processing throughput and energy efficiency for SPOGA, as evident from the results presented in the next section.

\section{Evaluation}
\subsection{Scalability Analysis}\label{sec:scalability}
Our scalability study is based on the methods for modeling photonic GEMM cores that are presented in prior works \cite{b1,b2,b12}. Using the modeling equations and parameters from \cite{b2}, we evaluated the maximum number of achievable dot products per GEMM core (denoted as M) and the maximum allowable vector size during dot product operations (denoted as N) for SPOGA and other architectures from prior works, as functions of operating data rate and input optical power. We considered input analog operand width of 4-bit, i.e., 2$^4$ analog levels. The results of our analysis are summarized in Table \ref{tab:NM}. As evident, SPOGA in general achieves the highest parallelism, i.e., the largest N$\times$M value. This is due to the full optical-analog dataflow in SPOGA, which saves optical power to keep it available in support of higher parallelism. 




\begin{figure}
\centerline{\includegraphics[width=18pc]{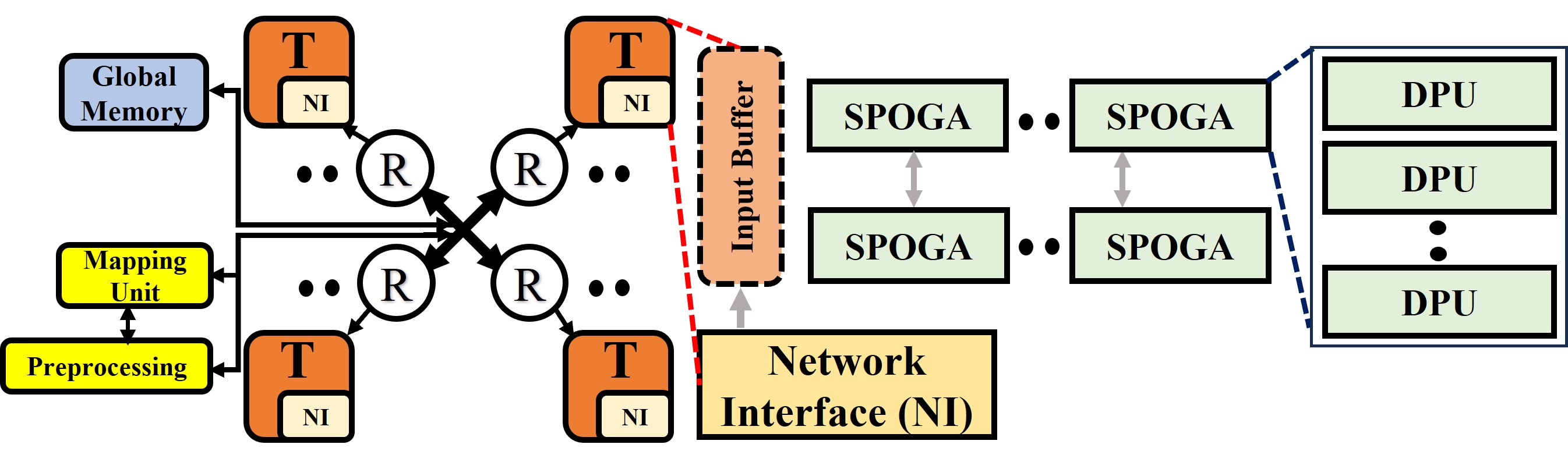}}
\caption{Schematic of system-level implementation of SPOGA.}\vspace*{-5pt}
\label{fig:syslvl}
\end{figure}

\begin{table}
    \centering
\caption{Results of Scalability Analysis. }
\label{tab:NM}
    \begin{tabular}{|m{2.2cm}|c|c|c|l|l|l|} \hline 
 & \multicolumn{6}{c|}{BR/DR}  \\ \hline 
         Architectures&  \multicolumn{2}{|c|}{1GS/s}& \multicolumn{2}{|c|}{5GS/s}& \multicolumn{2}{|c|}{10GS/s}\\ \hline 
 & N& M& N& M& N&M\\ \hline 
         HOLYLIGHT \cite{b3}&  43&  43& 21& 21& 15&15\\ \hline 
DEAPCNN \cite{b9}& 36& 36&17& 17& 12&12\\ \hline 
        MWA (1dBm)&  94&  16&  32& 16& 5&16\\ \hline
        MWA (5dBm)&  163&  16&  101& 16& 74&16\\ \hline
        MWA (10dBm)&  249&  16&  187& 16& 160&16\\ \hline
    \end{tabular}
\end{table}

\begin{table}[h]
\centering
\caption{Area and Power Overheads of ADC and DACs.}
\begin{tabular}{|c|l|l|l|}
\hline
\multicolumn{1}{|l|}{\textbf{Converters}} & \textbf{BRs (GS/s)} & \textbf{Area (mm$^2$)} & \textbf{Power (mw)} \\ \hline
\multirow{3}{*}{ADC} & 1 \cite{b13}  & 0.002   & 2.55 \\ \cline{2-4} 
                     & 5 \cite{b14}  & 0.021   & 11   \\ \cline{2-4} 
                     & 10 \cite{b15} & 0.103   & 29   \\ \hline
\multirow{3}{*}{DAC} & 1 \cite{b16}  & 0.00007 & 0.12 \\ \cline{2-4} 
                     & 5 \cite{b17}  & 0.06    & 26   \\ \cline{2-4} 
                     & 10 \cite{b18} & 0.06    & 30   \\ \hline
\end{tabular}
\label{tab:conv_val}
\end{table}


\subsection{System-Level Simulation Setup}
To assess the performance of our SPOGA accelerator, we employed a custom, transaction-level Python-based simulator. Within this simulator framework, we modeled SPOGA as depicted in Fig. \ref{fig:syslvl}. We used the modeling parameters, area/energy/latency parameters of hardware components, etc., from our prior works \cite{b1,b2}. The utilized parameter values for the ADCs and DACs are listed in Table \ref{tab:conv_val}. Subsequently, we conducted comprehensive simulations for the inference of four prominent CNN models: MobileNet V2, ShuffleNet V2, ResNet50, and GoogleNet. Throughout our evaluation, we focused on crucial metrics, including Frames-Per-Second (FPS), FPS/W (energy efficiency), and FPS/W/mm$^2$ (area efficiency). We conducted comparative assessments against two analog optical accelerators, namely MAW (HOLYLIGHT) \cite{b3} and AMW (DEAPCNN) \cite{b9}.

\subsection{Evaluation Results}

\begin{figure}
    \centering
    \subfigure[FPS (Log Scale)]
    {
        \includegraphics[width=\linewidth]{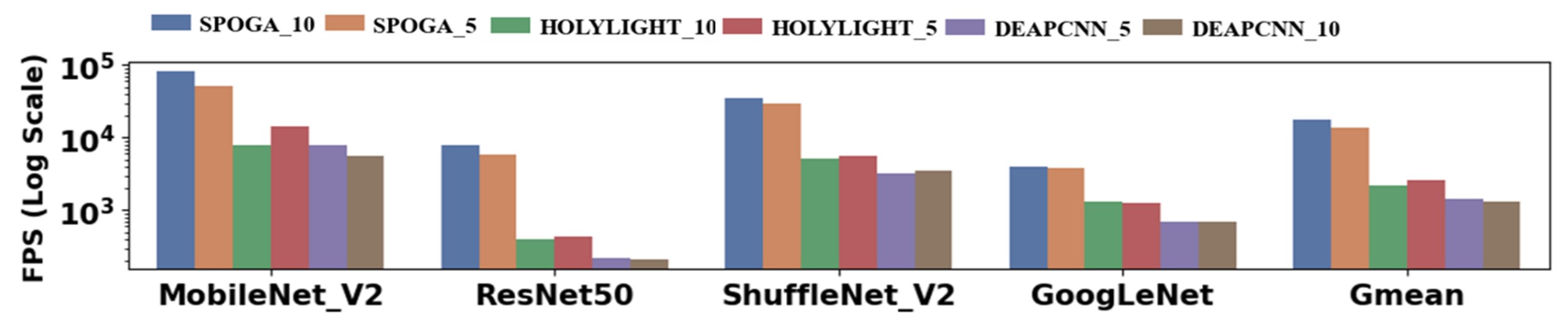}
        \label{fig:fps1}
    }
    \hfill
    \subfigure[FPS/W]
    {
        \includegraphics[width=\linewidth]{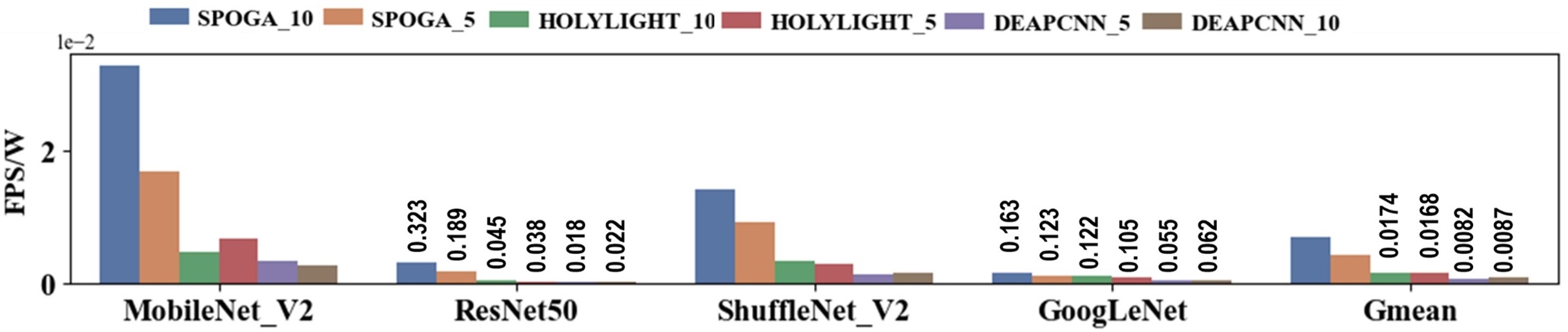}
        \label{fig:fps2}
    }
    \hfill
    \subfigure[FPS/W/Area]
    {
        \includegraphics[width=\linewidth]{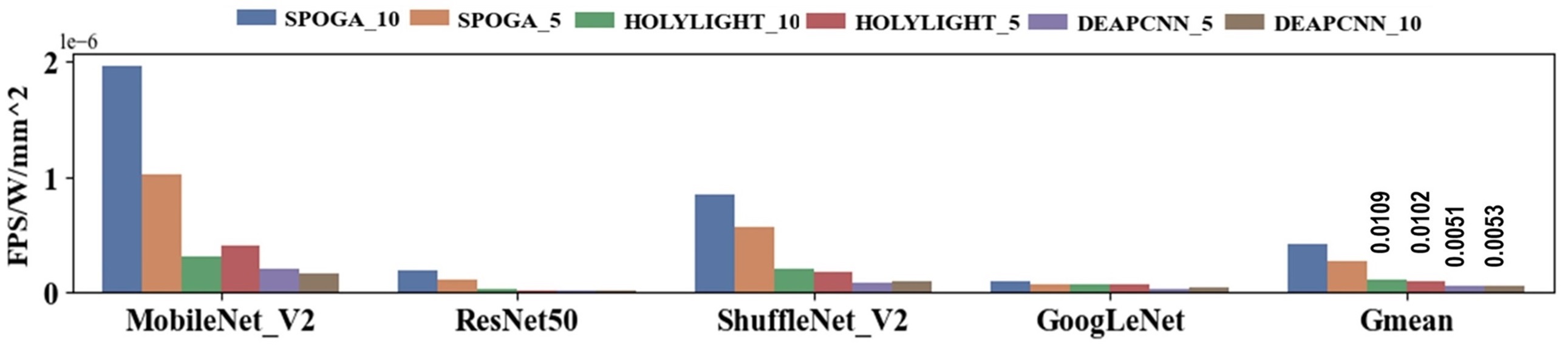}
        \label{fig:fps3}
    }
    \caption{Evaluation results for SPOGA versus HOLYLIGHT (MAW) and DEAPCNN (AMW) accelerators at 5 GS/s and 10 GS/s datarates.}
    \label{fig:fpsmetrics}
\end{figure}

Fig. \ref{fig:fps1} presents a logarithmic scale comparison of FPS results. Notably, $SPOGA \_10$ (SPOGA operating at 10 GS/s) exhibits a remarkable performance advantage, achieving 14.4$\times$ and 11.1$\times$ higher FPS than $DEAPCNN \_10$ and $HOLYLIGHT \_10$, respectively, as indicated by the geometric mean (gmean) values. 
Further, Fig. \ref{fig:fps2} and Fig. \ref{fig:fps3} present the FPS/W and FPS/W/Area results. Notably, $SPOGA \_10$ achieves 2$\times$ and 1.3$\times$ better FPS/W than $DEAPCNN \_10$ and $HOLYLIGHT \_10$ respectively. 
In addition, $SPOGA \_1$ shows 28.5$\times$ better FPS/W/Area compared to $DEAPCNN \_1$ and an impressive 22.2$\times$ better FPS/W/Area compared to $HOLYLIGHT \_1$. The better results obtained for SPOGA variants can be attributed to SPOGA's higher parallelism, extended optical-analog dataflow, and reduced overheads.

\section{Summary}
In this paper, we presented SPOGA, a photonic GEMM accelerator. SPOGA utilizes homodyne optical signals and in-transduction positional weighting of operands through an advanced optical-analog dataflow approach. The key advantages of SPOGA over traditional photonic and electronic GEMM accelerators are its ability to handle byte-size integer operands and its reduction in the overheads associated with bit-sliced integer arithmetic. These enable SPOGA to achieve significantly lower latency and higher throughput and energy efficiency compared to the state-of-the-art solutions. To validate the benefits of SPOGA, we evaluated its achievable parallelism, throughput, and energy efficiency, and compared the obtained results with other photonic GEMM accelerators. Our analysis shows that SPOGA supports higher parallelism and demonstrates up to 14.4$\times$ better throughput (FPS), at least 2$\times$ better energy efficiency per watt (FPS/Watt), and up to 28.5$\times$ better energy efficiency per square millimeter (FPS/Watt/mm²) than prior GEMM accelerators.

\bibliographystyle{IEEEtran}
\vspace*{-8pt}
\end{document}